\documentclass[twocolumn,aps,prd,amssymb,nofootinbib]{revtex4}
\usepackage{epsfig}
\usepackage{amsmath}
\usepackage{amssymb}

\begin{document}
\title{Palatini form of 1/R gravity}

\author{\'Eanna \'E.\ Flanagan}
\affiliation{Cornell University, Newman Laboratory, Ithaca, NY
14853-5001.}

\begin{abstract}
It has been suggested that the recent acceleration of the expansion of
the Universe is due a modified gravitational action consisting of the
Einstein-Hilbert term plus a term proportional to the reciprocal of
the Ricci scalar.   
Although the original version of this theory has been shown to be in
conflict with solar system observations, a modified Palatini version
of the theory, in which the metric and connection are treated as
independent variables in the variational principle, has been
suggested as a viable model of the cosmic acceleration.
We show that this theory is equivalent to a type of
scalar-tensor theory in which the scalar field kinetic energy
term in absent from the action, and in which the scalar field is
therefore not an independent dynamical degree of freedom.  
Integrating out the scalar field gives rise to 
additional interactions among the matter fields of the standard model
of particle physics at a
mass scale of order $10^{-3} \, {\rm eV}$ (the geometric mean of the
Hubble scale and the Planck scale), and so the theory is excluded by,
for example, electron-electron scattering experiments.   
\end{abstract}

\maketitle



The observed acceleration of the Universe's expansion
\cite{SN,wmap} is 
normally attributed to so-called dark energy, that is, an additional source
of gravity such as a cosmological constant or a quintessence field.  
However, it has recently been suggested that the acceleration is due
instead to a modification of gravity at cosmological distance scales
\cite{Capo,Carroll,Lue}.  In particular, Carroll {\it et. al.} \cite{Carroll}
suggested a gravitational action of the form
\begin{equation}
S = \frac{1}{2 \kappa^2} \int d^4 x \sqrt{-g} \left[ R -
  \frac{\mu^4}{R} \right],
\label{eq:carrollaction}
\end{equation}
where $R$ is the Ricci scalar, $\kappa^2 = 8 \pi G$ and $\mu$ is a
mass scale of order the Hubble scale; see also Refs.\
\cite{Vollick,Chiba,Cline,Meng,Odint1}.  Chiba \cite{Chiba} showed that 
the theory (\ref{eq:carrollaction}) is equivalent to a
scalar-tensor theory \cite{Damour92} with a very light scalar field
that couples to matter with gravitational strength.  This theory is
therefore ruled out by solar system experiments \cite{Chiba}.

However, a modified version of the theory (\ref{eq:carrollaction}) can be
obtained by treating the metric and the connection as independent
dynamical variables in the variational principle, as suggested by
Vollick \cite{Vollick}.  For general relativity, this Palatini or
first-order variational
principle is equivalent to the more usual variational principle where
the connection is taken to be determined by the metric.  However, for actions
that are nonlinear functions of the Ricci scalar, the Palatini
variational principle and the standard variational principle give rise
to inequivalent theories.  The Palatini version of the theory
(\ref{eq:carrollaction}), like the original version, can explain the
recent acceleration of the Universe's expansion \cite{Vollick,Meng},
but, unlike the original form, has not been shown to be in conflict
with solar system experiments.

In this paper we show that the Palatini form of the theory
(\ref{eq:carrollaction}) is equivalent to a type of 
scalar-tensor theory in which the scalar field kinetic energy
term in absent from the action.  Integrating out the scalar field
gives rise to additional interactions among the matter fields of the
standard model at a mass scale of order $10^{-3} \, {\rm eV}$,
and so the
theory is excluded by particle physics experiments.

We start by reviewing the equivalence of 
higher-order gravity theories of the form
(\ref{eq:carrollaction}) to scalar-tensor theories
\cite{TT,Wands}.  
Consider an action of the form
\begin{equation}
S[{\bar g}_{\mu\nu},\psi_{\rm m}] = \frac{1}{2 \kappa^2} \int d^4 x
\sqrt{- {\bar g}} f({\bar R}) + S_{\rm m}[{\bar g}_{\mu\nu},\psi_{\rm
    m}].
\label{eq:action0}
\end{equation}
Here ${\bar g}_{\mu\nu}$ is the metric (which we
have barred for later 
notational convenience), ${\bar R}$ is its Ricci scalar, and $f$ is
any function.  The second term is the matter action $S_{\rm m}$, which
is some functional of the matter fields $\psi_{\rm m}$ and of the
metric.  We use units in which $\hbar=c=1$, and we use the sign
conventions of Ref.\ \cite{MTW}.  

We introduce an extra scalar field $\varphi$ into the theory by
defining a modified action \cite{Wands}
\begin{eqnarray}
{\tilde S}[{\bar g}_{\mu\nu},\varphi,\psi_{\rm m}] &=& 
\frac{1}{2 \kappa^2} \int d^4 x
\sqrt{- {\bar g}} \left[ f(\varphi) + ({\bar R} - \varphi)
  f'(\varphi)\right] \nonumber \\
&& + S_{\rm m}[{\bar g}_{\mu\nu},\psi_{\rm m}].
\label{eq:action1}
\end{eqnarray}
The $\varphi$ equation of motion obtained from this action is $\varphi
= {\bar R}$, as long as $f''(\varphi)\ne0$, and thus the theory
(\ref{eq:action1}) is classically equivalent to the
original theory (\ref{eq:action0}).  Next, we define a conformally
rescaled metric $g_{\mu\nu}$ by
\begin{equation}
{\bar g}_{\mu\nu} = e^{2 \alpha(\varphi)} g_{\mu\nu},
\label{eq:ctransform}
\end{equation}
where $\alpha$ is the function of $\varphi$ given by
\begin{equation}
e^{2 \alpha(\varphi)} f'(\varphi) =1,
\label{eq:fdef}
\end{equation}
and we introduce the canonically normalized scalar field
\begin{equation}
\Phi = - \frac{\sqrt{6}}{\kappa} \alpha(\varphi) = \frac{\sqrt{6}}{2
  \kappa} \ln f'(\varphi).
\label{eq:Phidef}
\end{equation}
The action now simplifies to 
\begin{eqnarray}
{\tilde S}[g_{\mu\nu},\Phi,\psi_{\rm m}] &=& 
\int d^4 x
\sqrt{- g} \bigg[ \frac{R}{2 \kappa^2} -\frac{1}{2} (\nabla \Phi)^2 - V(\Phi)
\bigg] \nonumber \\
&& + S_{\rm m}[e^{2 \alpha(\Phi)} g_{\mu\nu},\psi_{\rm m}].
\label{eq:action3}
\end{eqnarray}
This has the form of a scalar-tensor theory \cite{Damour92}, written in
terms of the 
Einstein-frame metric $g_{\mu\nu}$.  The potential for the scalar
field is 
\begin{equation}
V = \frac{\varphi f'(\varphi) - f(\varphi)}{2 \kappa^2 f'(\varphi)^2},
\label{eq:pot}
\end{equation}
and the coupling function $\alpha(\Phi)$ is given by
\begin{equation}
\alpha(\Phi) = - \frac{\kappa}{\sqrt{6}} \Phi.
\label{eq:alphadef}
\end{equation}

In this class of theories, the scalar field couples to matter with essentially
the same strength as does gravity \cite{Chiba}.  A measure of the ratio of the
scalar coupling to the gravitational coupling is \cite{Damour92}
\begin{equation}
1 - \gamma  = \frac{4 (d\alpha /d\Phi)^2 / \kappa^2}{1 + 2
  (d\alpha/d\Phi)^2 / \kappa^2} = \frac{1}{2},
\end{equation}
where $\gamma$ is the PPN parameter.  Solar system VLBI experiments
show that $|\gamma -1| \le 3 \times 10^{-4}$ \cite{Will}, and thus the
theory (\ref{eq:action3}) is ruled out unless the potential $V(\Phi)$
is such that the field $\Phi$ is massive and short ranged
\cite{Chiba}.\footnote{Note that this conclusion is independent of the choice
of the function $f({\bar R})$ in the original action (\ref{eq:action0}),
contrary to the claims in Refs.\ \protect{\cite{Cline}}.}
For the Carroll {\it et. al.} model (\ref{eq:carrollaction}), the
potential (\ref{eq:pot}) is \cite{Carroll,Chiba}
\begin{equation}
V(\Phi) = \frac{\mu^2}{\kappa^2} \exp \left[ - 2 \sqrt{ \frac{2}{3}}
  \kappa \Phi \right] \sqrt{ \exp \left[ \sqrt{ \frac{2}{3}}  \kappa
  \Phi \right] -1}.
\label{eq:pot1}
\end{equation}
For this theory to explain the cosmic acceleration, we require $\mu \sim
H_0$, where $H_0 \sim 1.5 \times 10^{-33} \, {\rm eV}$ is the Hubble
scale, and the resulting present-day effective mass $\sqrt{V''(\Phi)}$
of the scalar field is $\sim H_0$ at $\kappa \Phi \sim 1$.
Thus this theory is not viable \cite{Chiba}.

Turn now to the Palatini form of the theory
(\ref{eq:action0}).  We first review the formulation of this theory
given by Vollick \cite{Vollick}.  The action is a function of the Jordan-frame
metric ${\bar g}_{\mu\nu}$, a symmetric connection ${\hat \nabla}_\mu$,
and the matter fields $\psi_{\rm m}$.  We can use instead of the
connection ${\hat \nabla}_\mu$ the tensor field $H^\lambda_{\mu\nu}$
defined by
$
{\hat \nabla}_\mu v^\lambda - {\bar \nabla}_\mu v^\lambda =
H^\lambda_{\mu\nu} v^\nu
$
for any vector field $v^\lambda$, where ${\bar \nabla}_\mu$ is the
connection determined by the metric ${\bar g}_{\mu\nu}$.  The action takes
the form \cite{Vollick}
\begin{equation}
S[{\bar g}_{\mu\nu},H^\mu_{\nu\lambda},\psi_{\rm m}] = \frac{1}{2
  \kappa^2} \int d^4 x \sqrt{- {\bar g}} f({\hat R}) +
S_{\rm m}[{\bar g}_{\mu\nu},\psi_{\rm m}],
\label{eq:action4}
\end{equation}
where ${\hat R} = {\bar g}^{\mu\nu} \,
{\hat R}_{\mu\nu}$ and ${\hat R}_{\mu\nu}$ is the Ricci
tensor of the connection ${\hat \nabla}_\mu$, given by
\begin{eqnarray}
{\hat R}_{\mu\nu} = {\bar R}_{\mu\nu} + {\bar \nabla}_\lambda
H^\lambda_{\mu\nu} - {\bar \nabla}_\mu H^\lambda_{\lambda\nu}
+ H^\lambda_{\lambda\sigma} H^\sigma_{\mu\nu} - H^\lambda_{\mu\sigma}
H^\sigma_{\lambda\nu}.
\label{eq:ricci}
\end{eqnarray}
Varying the action with respect to $H^\lambda_{\mu\nu}$ gives an
equation of motion whose unique solution is \cite{Vollick}
\begin{equation}
H^\lambda_{\mu\nu} = \delta^\lambda_{(\mu} {\bar \nabla}_{\nu)} \ln
f'({\hat R}) - \frac{1}{2} {\bar g}_{\mu\nu} {\bar g}^{\lambda\sigma}
{\bar \nabla}_\sigma \ln f'({\hat R}).
\label{eq:sol1}
\end{equation}
Equation (\ref{eq:sol1}) is equivalent to the statement that the
connection ${\hat \nabla}_\mu$ is compatible with the metric
$
f'({\hat R}) {\bar g}_{\mu\nu}.
$
Varying the action with respect to the
metric yields
\begin{equation}
f'({\hat R}) {\hat R}_{\mu\nu} - \frac{1}{2} f({\hat R}) {\bar
  g}_{\mu\nu} = \kappa^2 {\bar T}_{\mu\nu},
\label{eq:eom2}
\end{equation}
where ${\bar T}_{\mu\nu}$ is the stress-energy tensor.
The trace of this equation gives the algebraic relation
\begin{equation}
f'({\hat R}) {\hat R} - 2 f({\hat R}) = \kappa^2 {\bar T},
\label{eq:eom3}
\end{equation}
which can be used to solve for ${\hat R}$ in terms of ${\bar T} =
{\bar g}^{\mu\nu} {\bar T}_{\mu\nu}$.  The Ricci scalar of the metric ${\bar
  g}_{\mu\nu}$ is then given by the contraction of Eq.\ (\ref{eq:ricci}):
\begin{equation}
{\bar R} = {\hat R} + 3 {\bar \Box} \ln f'({\hat R}) + \frac{3}{2} [ {\bar
  \nabla} \ln f'({\hat R}) ]^2.
\label{eq:eom4}
\end{equation}
The traceless part of the equation of motion (\ref{eq:eom2}) 
can be written in terms of the Ricci tensor ${\bar R}_{\mu\nu}$ as
\begin{eqnarray}
&& \bigg[ {\bar R}_{\mu\nu} - {\bar \nabla}_\mu {\bar \nabla}_\nu \ln
  f'({\hat R}) + \frac{1}{2} {\bar \nabla}_\mu \ln f'({\hat R}) {\bar
  \nabla}_\nu \ln f'({\hat R})\bigg]_{\rm TL} \nonumber \\
\mbox{} && = \frac{\kappa^2}{f'({\hat R})} \left[ T_{\mu\nu} \right]_{\rm TL},
\label{eq:eom5}
\end{eqnarray}
where TL means ``the traceless part of''.  Equations (\ref{eq:eom3})
-- (\ref{eq:eom5}) replace the usual 
Einstein equations, and the corresponding Friedmann-Robertson-Walker
cosmological models have been studied by Refs.\ \cite{Vollick,Meng}.

We now show that the theory (\ref{eq:action4}) is equivalent to a type
of scalar tensor theory.  
First, note that the connection is always on-shell compatible with
some conformal transform $e^{2 \chi} {\bar g}_{\mu\nu}$ of the metric
${\bar g}$, so that 
\begin{equation}
H^\lambda_{\mu\nu} = \delta^\lambda_{(\mu} {\bar \nabla}_{\nu)} \ln
\chi - \frac{1}{2} {\bar g}_{\mu\nu} {\bar g}^{\lambda\sigma}
{\bar \nabla}_\sigma \ln \chi
\label{eq:replace}
\end{equation}
for some scalar field $\chi$.  Substituting this ansatz into the
action (\ref{eq:action4}) gives
\begin{eqnarray}
S[{\bar g}_{\mu\nu},\chi,\psi_{\rm m}] &=& \frac{1}{2
  \kappa^2} \int d^4 x \sqrt{- {\bar g}} \, f[{\bar R} - 6 ({\bar
  \nabla}   \chi)^2 - 6 {\bar \Box} \chi] \nonumber \\
&& + S_{\rm m}[{\bar g}_{\mu\nu},\psi_{\rm m}].
\label{eq:action6}
\end{eqnarray}
All stationary points of the action (\ref{eq:action4}) will also be
stationary points
of the action (\ref{eq:action6}), by construction.  However, the
converse is not true, since solutions of the equations of motion
of the theory (\ref{eq:action6})
are stationary only under a restricted set of
variations of $\delta H^\mu_{\nu\lambda}$ of the connection.  We
discuss this point further below.

We next use the technique discussed above to eliminate the nonlinear
kinetic terms by introducing an auxiliary scalar field $\varphi$.  The
modified action is
\begin{eqnarray}
\label{eq:action7}
&&{\tilde S}[{\bar g}_{\mu\nu},\chi,\varphi,\psi_{\rm m}] = \frac{1}{2
  \kappa^2} \int d^4 x \sqrt{- {\bar g}} \bigg\{ f(\varphi) 
  \\
\nonumber
\mbox{} && + \left[ {\bar R} - 6 ( {\bar \nabla} \chi)^2 - 6 {\bar
  \Box} \chi - \varphi \right] f'(\varphi) \bigg\}
+ S_{\rm m}[{\bar g}_{\mu\nu},\psi_{\rm m}].
\nonumber
\end{eqnarray}
As before the $\varphi$ equation of motion is 
\begin{equation}
\varphi = {\bar R} - 6 ( {\bar \nabla} \chi)^2 - 6 {\bar
  \Box} \chi, 
\label{eq:varphieqn}
\end{equation}
so the theories (\ref{eq:action6}) and
(\ref{eq:action7}) are classically equivalent.  We next transform
to the Einstein conformal frame $g_{\mu\nu}$ using the relations 
(\ref{eq:ctransform}) and (\ref{eq:fdef}).  We also define the canonically
normalized version $\Phi$ of the field $\varphi$ by Eq.\ (\ref{eq:Phidef}) as
before, and we diagonalize the kinetic energy term in the action by
using instead of the field $\chi$ the quantity
\begin{equation}
\Psi = \sqrt{6} [\alpha(\varphi) + \chi]/\kappa.
\label{eq:Psidef}
\end{equation}
The resulting action is
\begin{eqnarray}
{\tilde S}[g_{\mu\nu},\Phi,\Psi,\psi_{\rm m}] &=& 
\int d^4 x
\sqrt{- g} \bigg[ \frac{R}{2 \kappa^2} -\frac{1}{2} (\nabla \Psi)^2 - V(\Phi)
\bigg] \nonumber \\
&& + S_{\rm m}[e^{2 \alpha(\Phi)} g_{\mu\nu},\psi_{\rm m}],
\label{eq:action8}
\end{eqnarray}
where the potential $V(\Phi)$ and coupling function $\alpha(\Phi)$ are
given by the same expressions (\ref{eq:pot}) [or (\ref{eq:pot1})] and
(\ref{eq:alphadef}) as before.  

We now address the issue of spurious solutions of the equations of
motion of the theory (\ref{eq:action8}) that do not satisfy the
equations of motion of the original theory (\ref{eq:action4}).   
We show that the appropriate subclass of solutions of the theory
(\ref{eq:action8}) are those solutions with $\Psi = {\rm const}$.  
To see this, note that for solutions of the original theory
(\ref{eq:action4}), the connection is compatible with the metric
$f'({\hat R}) {\bar g}_{\mu\nu}$, by Eq.\ (\ref{eq:sol1}).    
Using Eqs.\ (\ref{eq:varphieqn}), (\ref{eq:fdef}) and (\ref{eq:Psidef})
this metric can be written as
\begin{equation}
f'(\varphi) {\bar g}_{\mu\nu} = e^{-2 \alpha} {\bar g}_{\mu\nu} = e^{-
  2 \kappa \Psi/\sqrt{6}} \left( e^{2 \chi} {\bar g}_{\mu\nu} \right).
\end{equation}
Comparing with the ansatz (\ref{eq:replace}) shows that on-shell we
must have $\Psi = $ constant, as claimed.  It is also possible to show
that this is the only restriction on solutions.

Therefore, the final action can be obtained simply by deleting the
field $\Psi$ from the action (\ref{eq:action8}):
\begin{eqnarray}
{\tilde S}[g_{\mu\nu},\Phi,\psi_{\rm m}] &=& 
\int d^4 x
\sqrt{- g} \bigg[ \frac{R}{2 \kappa^2}  - V(\Phi)
\bigg] \nonumber \\
&& + S_{\rm m}[e^{2 \alpha(\Phi)} g_{\mu\nu},\psi_{\rm m}].
\label{eq:action9}
\end{eqnarray}
This action has exactly the same form as the scalar-tensor form
(\ref{eq:action3}) of the standard-variation version of the theory,
except that the 
kinetic energy term for the field $\Phi$ has been deleted.
Thus, somewhat paradoxically, allowing the connection to be an
independent dynamical variable has reduced rather than increased the
number of degrees of freedom of the theory; the field $\Phi$ is no
longer dynamical but is determined from an algebraic equation (see below).

The equations of motion of this formulation of the theory are
significantly simpler than those of the original formulation.  They are
\begin{equation}
\frac{1}{\kappa^2} G_{\mu\nu} = -V(\Phi) g_{\mu\nu} + e^{2
  \alpha(\Phi)} {\bar T}_{\mu\nu},
\label{eq:eom1a}
\end{equation}
and
\begin{equation}
V'(\Phi) = \alpha'(\Phi) e^{4 \alpha(\Phi)} {\bar T},
\label{eq:eom2a}
\end{equation}
where ${\bar T}_{\mu\nu} = - (2/\sqrt{-{\bar g}}) \delta S_{\rm m} /
\delta {\bar  
g}^{\mu\nu}$ is the Jordan-frame stress energy tensor and ${\bar T} = {\bar
g}^{\mu\nu} {\bar T}_{\mu\nu}$.  Note that Eq.\ (\ref{eq:eom2a})
implies that $\Phi$ is an algebraic function of ${\bar T}$.  
It can be interpreted as saying that the field always sits at the
extremum point of the matter-corrected potential
$
V(\Phi) - e^{4 \alpha(\Phi)} {\bar T}/4.
$

We next discuss the nature of the solutions of the algebraic equation
(\ref{eq:eom2a}) for $\Phi$.  
We restrict attention to the model (\ref{eq:pot1}) of Carroll {\it
  et. al.}, and to the case of negative ${\bar T}$.  We define
$\rho_{\rm c} = \mu^2 /\kappa^2$, which is a critical energy density of
order the present cosmological energy density.  
When $|{\bar T}| \ll \rho_c$,
we have
$\Phi \approx \Phi_{\rm max} = \sqrt{3/2} \ln(4/3)/\kappa$ (where
$\Phi_{\rm max}$ is the value
of $\Phi$ at the local maximum of the potential), $V(\Phi) \approx
V(\Phi_{\rm max}) = 3 \sqrt{3} \rho_c / 16$, and $e^{2 \alpha(\Phi)}
\approx 3/4$. 
When $|{\bar T}| \gg \rho_c$, on the other hand, the solution is 
$\kappa \Phi \sim (\rho_{\rm c} / |{\bar T}|)^2 \sqrt{3/2} \ll 1$,
corresponding to $V(\Phi) \sim \rho_c^2 / |{\bar T}|$ and $e^{2
  \alpha(\Phi)} \sim 1$.  Thus, up to fractional corrections of order
$\rho_c^2 / |{\bar T}|^2$, the equation of motion (\ref{eq:eom1a})
reduces to Einstein's equation.

However, in terrestrial and astrophysical environments, these
conclusions are only valid when the stress energy 
tensor ${\bar T}_{\mu\nu}$ is interpreted to be the true, microscopic
stress energy, and not the
macroscopic, spatially-averaged stress energy tensor.
What this means is as follows.  The
true, microscopic stress energy tensor ${\bar T}_{\mu\nu}$ will vary
over atomic scales in matter, from $|{\bar T}| \gg \rho_c$ inside
atoms, to $|{\bar T}| \ll \rho_c$ in between atoms.  Let us write
${\bar T}_{\mu\nu} = \langle {\bar T}_{\mu\nu} \rangle + \delta {\bar
  T}_{\mu\nu}$, $g_{\mu\nu} = \langle g_{\mu\nu} \rangle + \delta
g_{\mu\nu}$, where $\langle {\bar T}_{\mu\nu} \rangle $ and $\langle
g_{\mu\nu} 
\rangle$ are the
spatial averages of the stress energy tensor and the metric over some
lengthscale large compared to atomic scales, and $\delta T_{\mu\nu}$
and $\delta g_{\mu\nu}$ are the fluctuations due to the microscopic
structure of matter.  For general relativity, we have $\delta
g_{\mu\nu} \ll 1$, and so the fluctuations can be treated as a linear
perturbation in Einstein's equation.  This guarantees that Einstein's
equation continues to hold to a good approximation with ${\bar T}_{\mu\nu}$
replaced by $\langle {\bar T}_{\mu\nu} \rangle$ and with $g_{\mu\nu}$
replaced by $\langle g_{\mu\nu} \rangle$.

However, this property does not hold for the theory of gravity given
by Eqs.\ (\ref{eq:eom1a}) and (\ref{eq:eom2a}).  If we write
$
\Phi = \langle \Phi \rangle + \delta \Phi,
$
then since the energy density varies from scales $\gg \rho_c$ inside
atoms to scales $\ll \rho_c$ in between, it follows that $\kappa \delta
\Phi$ is of order unity.  Hence, $\delta \Phi$ cannot be treated as a
linear perturbation in Eq.\ (\ref{eq:eom1a}); for example one
cannot make the replacement 
$
\langle e^{2 \alpha(\Phi)} \rangle  = e^{2 \alpha(\langle
  \Phi\rangle)}.
$
Thus Eq.\ (\ref{eq:eom1a}) and (\ref{eq:eom2a}) are not valid
with the fields ${\bar T}_{\mu\nu}$, $g_{\mu\nu}$ and $\Phi$ all
replaced by their spatial averages.  In particular, this invalidates
the FRW cosmological models obtained from Eqs.\ (\ref{eq:eom1a}) --
(\ref{eq:eom2a}) [or equivalently Eqs.\ (\ref{eq:eom3}) --
  (\ref{eq:eom5})] with ${\bar T}_{\mu\nu}$ taken to be that of
pressureless matter \cite{Vollick,Meng}.    

The strong dependence of solutions of the equations of motion on the
microphysical structure of matter suggests that one should treat the
matter source in terms of quantum field theory.  In this context, one can 
integrate out the field $\Phi$ by solving its equation of motion and
backsubstituting into the action (\ref{eq:action9}).  One then finds that 
additional interactions are generated among the various matter fields
of the standard model of particle physics that are sufficiently large
to be in severe violation of experimental bounds.  

We illustrate this effect by taking the matter action to be the Dirac
action for free electrons:
\begin{equation}
S_{\rm m}[{\bar g}_{\mu\nu}, \Psi_{\rm e}] = \int d^4 x \sqrt{- {\bar g}}
\, {\bar \Psi}_{\rm e} \left[ i {\bar \gamma}^\mu \nabla_\mu - m_{\rm e}
  \right] \Psi_{\rm e}.
\end{equation}
Here $\Psi_{\rm e}$ is a Dirac spinor and $m_{\rm e}$ is the electron
mass, and ${\bar \gamma}_\mu$ are the 
Dirac matrices associated with the metric ${\bar g}_{\mu\nu}$,
satisfying
$
{\bar \gamma}^\mu {\bar \gamma}^\nu + {\bar \gamma}^\nu {\bar
  \gamma}^\mu = - 2 {\bar g}^{\mu\nu}.
$
Substituting into Eq.\ (\ref{eq:action9}) gives for the total action
\begin{eqnarray}
&& {\tilde S}[g_{\mu\nu},\Phi,\Psi_{\rm e}] =
\int d^4 x
\sqrt{- g} \bigg[ \frac{R}{2 \kappa^2}  - V(\Phi)
\nonumber \\
&& 
+ i e^{3 \alpha(\Phi)} {\bar \Psi}_{\rm e} \gamma^\mu \nabla_\mu  
 \Psi_{\rm e} - e^{4 \alpha(\Phi) } m_{\rm e} {\bar \Psi}_{\rm e} \Psi_{\rm e}
\bigg],
\label{eq:action10}
\end{eqnarray}
where $\gamma^\mu = e^{-\alpha} {\bar \gamma}^\mu$ are the Dirac
matrices associated with the Einstein frame metric $g_{\mu\nu}$.  
The equation of motion for $\Phi$ is
$V'(\Phi) = 3 \alpha'(\Phi) e^{3 \alpha(\Phi)} i {\bar \Psi}_{\rm e}
 \gamma^\mu \nabla_\mu \Psi_{\rm e} 
- 4 \alpha'(\Phi) e^{4
 \alpha(\Phi)} m_{\rm e} {\bar \Psi}_{\rm e} \Psi_{\rm e}$.
Solving this equation for $\Phi$ using Eqs.\ 
(\ref{eq:alphadef}) and (\ref{eq:pot1}) yields
\begin{equation}
\kappa \Phi = \kappa \Phi_{\rm max} - \frac{1}{\sqrt{2}} {\cal M} -
\sqrt{\frac{3}{8}} {\cal K} + O({\cal K}^2, {\cal M}^2, {\cal K} {\cal M}),
\label{eq:Phisoln}
\end{equation}
where ${\cal K}$ and ${\cal M}$ are the dimensionless
quantities 
${\cal K} = i \kappa^2{\bar
 \Psi}_{\rm e} \gamma^\mu \nabla_\mu  \Psi_{\rm e} / \mu^2$,
$ {\cal M} = \kappa^2 m_{\rm e} {\bar
 \Psi}_{\rm e} \Psi_{\rm e} / \mu^2$.
Substituting the solution (\ref{eq:Phisoln}) back into the action
(\ref{eq:action10}), and performing the constant conformal
transformation $g_{\mu\nu} \to (4/3) g_{\mu\nu}$ gives   
\begin{eqnarray}
&& {\tilde S}[g_{\mu\nu},\Psi_{\rm e}] =
\int d^4 x
\sqrt{- g} \bigg[ \frac{R}{2 {\tilde \kappa}^2}  - \Lambda 
+ i {\bar \Psi}_{\rm e} \gamma^\mu \nabla_\mu \Psi_{\rm e} 
\nonumber \\
&& 
- m_{\rm e} {\bar \Psi}_{\rm e} \Psi_{\rm e} - \frac{3 \sqrt{3}}{16
  m_*^4} (i {\bar \Psi}_{\rm e} \gamma^\mu \nabla_\mu  \Psi_{\rm e})^2
- \frac{1}{\sqrt{3}} \frac{m_{\rm e}^2}{m_*^4} ( {\bar
 \Psi}_{\rm e} \Psi_{\rm e})^2
\nonumber \\
&& 
+ \sqrt{\frac{3}{4}} \frac{m_{\rm e}} {m_*^4} (i {\bar \Psi}_{\rm e} \gamma^\mu
\nabla_\mu \Psi_{\rm e}) ( {\bar
 \Psi}_{\rm e} \Psi_{\rm e}) + \ldots
\bigg],
\label{eq:action11}
\end{eqnarray}
where ${\tilde \kappa} = \sqrt{4/3} \kappa$, $m_* = \sqrt{\mu /
\kappa}$ and $\Lambda = \mu^2 / (\sqrt{3} \kappa^2)$ is the
induced cosmological constant.

The last three terms in the action (\ref{eq:action11}) are corrections
to the standard model.  These corrections 
are characterized by the mass scale $m_*$, which is roughly the geometric mean
of the Planck and the Hubble scales, of order $10^{-3} \, {\rm eV}$.
Since this mass scale is so small, it is clear that the
action (\ref{eq:action11}) is in severe conflict with particle physics
experiments, for example electron-electron scattering.  Thus the
original gravitation theory (\ref{eq:action4}) is ruled out. 

A simple physical explanation for this effect is the following.
Consider a region of space where the
energy density $\rho$ is of order $\rho_c$ and which varies over a
lengthscale ${\cal L} \ll H_0^{-1}$, where $H_0$ is the Hubble scale.
Then, the corresponding Einstein-frame gravitational field will be
negligible, but the quantity $e^{2 \alpha}$ and the perturbation to the
Jordan-frame metric will be of order unity, from Eq.\
(\ref{eq:eom2a}).  Therefore the gravitational acceleration
experienced by test particles will be $\sim c^2 / {\cal L}$, of
order the gravitational acceleration produced by a black hole of size
$\sim {\cal L}$ near its horizon.  Such gravitational accelerations
are significantly 
larger than in general relativity.  Although the theory
(\ref{eq:action4}) was designed to
produce deviations from general relativity only at very large
lengthscales $\sim \mu^{-1} \sim H_0^{-1}$, in fact deviations from
general relativity can be manifest at much smaller distance scales, as
long as the local radius of curvature of spacetime $\sim \sqrt{c^3 /
  (G \rho)}$ is of order $\mu^{-1}$, and the lengthscale ${\cal L}$
over which the density varies satisfies ${\cal L} \ll \mu^{-1}$.
Since the density scale $\rho \sim \rho_c$ can be achieved in the
scattering of sufficiently low energy elementary particles, such
particles experience the non-conventional large gravitational forces
discussed above, and the particle scattering cross-sections are
therefore affected.

\medskip

I thank Nima Arkani-Hamed and Jim Cline for helpful discussions, 
and the high energy theory group at Harvard University, where this
paper was written, for its hospitality.  
This research was supported in part by NSF grant PHY-0140209.

\end{document}